\documentclass[preprint, superscriptaddress,showpacs,amsmath,amssymb]{revtex4}

\usepackage{graphicx}

\begin{document}
\title{ARPES studies of cuprate Fermiology: superconductivity, pseudogap, and quasiparticle dynamics}
\author{I. M. Vishik}
\affiliation {Stanford Institute for Materials and Energy Sciences, SLAC National Accelerator Laboratory, 2575 Sand Hill Road, Menlo Park, CA 94025, USA}
\affiliation{Geballe Laboratory for Advanced Materials, Departments of Physics and Applied Physics, Stanford University, Stanford, CA 94305, USA}
\author{W. S. Lee}
\affiliation {Stanford Institute for Materials and Energy Sciences, SLAC National Accelerator Laboratory, 2575 Sand Hill Road, Menlo Park, CA 94025, USA}
\affiliation{Geballe Laboratory for Advanced Materials, Departments of Physics and Applied Physics, Stanford University, Stanford, CA 94305, USA}
\author{R.-H. He}
\affiliation {Stanford Institute for Materials and Energy Sciences, SLAC National Accelerator Laboratory, 2575 Sand Hill Road, Menlo Park, CA 94025, USA}
\affiliation{Geballe Laboratory for Advanced Materials, Departments of Physics and Applied Physics, Stanford University, Stanford, CA 94305, USA}
\author{M. Hashimoto}
\affiliation {Stanford Institute for Materials and Energy Sciences, SLAC National Accelerator Laboratory, 2575 Sand Hill Road, Menlo Park, CA 94025, USA}
\affiliation{Geballe Laboratory for Advanced Materials, Departments of Physics and Applied Physics, Stanford University, Stanford, CA 94305, USA}
\affiliation{Advanced Light Source, Lawrence Berkeley National Lab, Berkeley, CA 94720, USA}
\author{T. P. Devereaux}
\affiliation {Stanford Institute for Materials and Energy Sciences, SLAC National Accelerator Laboratory, 2575 Sand Hill Road, Menlo Park, CA 94025, USA}
\author{Z.-X. Shen}
\affiliation {Stanford Institute for Materials and Energy Sciences, SLAC National Accelerator Laboratory, 2575 Sand Hill Road, Menlo Park, CA 94025, USA}
\affiliation{Geballe Laboratory for Advanced Materials, Departments of Physics and Applied Physics, Stanford University, Stanford, CA 94305, USA}


\date{\today}

\begin{abstract}
We present angle-resolved photoemission spectroscopy (ARPES) studies of the cuprate high-temperature superconductors which elucidate the relation between superconductivity and the pseudogap and highlight low-energy quasiparticle dynamics in the superconducting state.  Our experiments suggest that the pseudogap and superconducting gap represent distinct states, which coexist below T$_c$.  Studies on Bi-2212 demonstrate that the near-nodal and near-antinodal regions behave differently as a function of temperature and doping, implying that different orders dominate in different momentum-space regions.  However, the ubiquity of sharp quasiparticles all around the Fermi surface in Bi-2212 indicates that superconductivity extends into the momentum-space region dominated by the pseudogap, revealing subtlety in this dichotomy. In Bi-2201, the temperature dependence of antinodal spectra reveals particle-hole asymmetry and anomalous spectral broadening, which may constrain the explanation for the pseudogap.  Recognizing that electron-boson coupling is an important aspect of cuprate physics, we close with a discussion of the multiple 'kinks' in the nodal dispersion.  Understanding these may be important to establishing which excitations are important to superconductivity.
\end{abstract}

\maketitle
\section{introduction}
Two general approaches are commonly taken to understand unconventional superconductivity, such as that in the high-temperature superconducting cuprates.  The first is to study the state which exists at temperatures higher than the superconducting transition temperature, T$_c$, in order to posit how it may become unstable to superconductivity.  The second is to search for bosonic excitations which might bind electrons into Cooper pairs.  Both approaches rely on examining the quasiparticle properties or Fermiology of the charge carriers.  Angle-resolved photoemission spectroscopy (ARPES) is an ideal tool for studying cuprate Fermiology because it directly measures the occupied part of the single-particle spectral function in momentum space.\cite{ARPES_Review}  ARPES has been instrumental in revealing key properties of the cuprates including: a \textit{d}-wave superconducting gap,\cite{Shen:dwave,Ding:ARPES_SC_gap} the pseudogap state above T$_c$,\cite{Loeser:NormalState,DestructionFS_Norman} and ubiquitous electron-boson coupling.\cite{ElectronPhononCoupling:Lanzara,KinkPRL:Bogdanov,Zhou:EPhCoupling,Cuk:ReviewEPh_coupling,Johnston:ReviewRenormalization}

\begin{figure} [t]
\includegraphics [type=jpg,ext=.jpg,read=.jpg,clip, width=3.7 in]{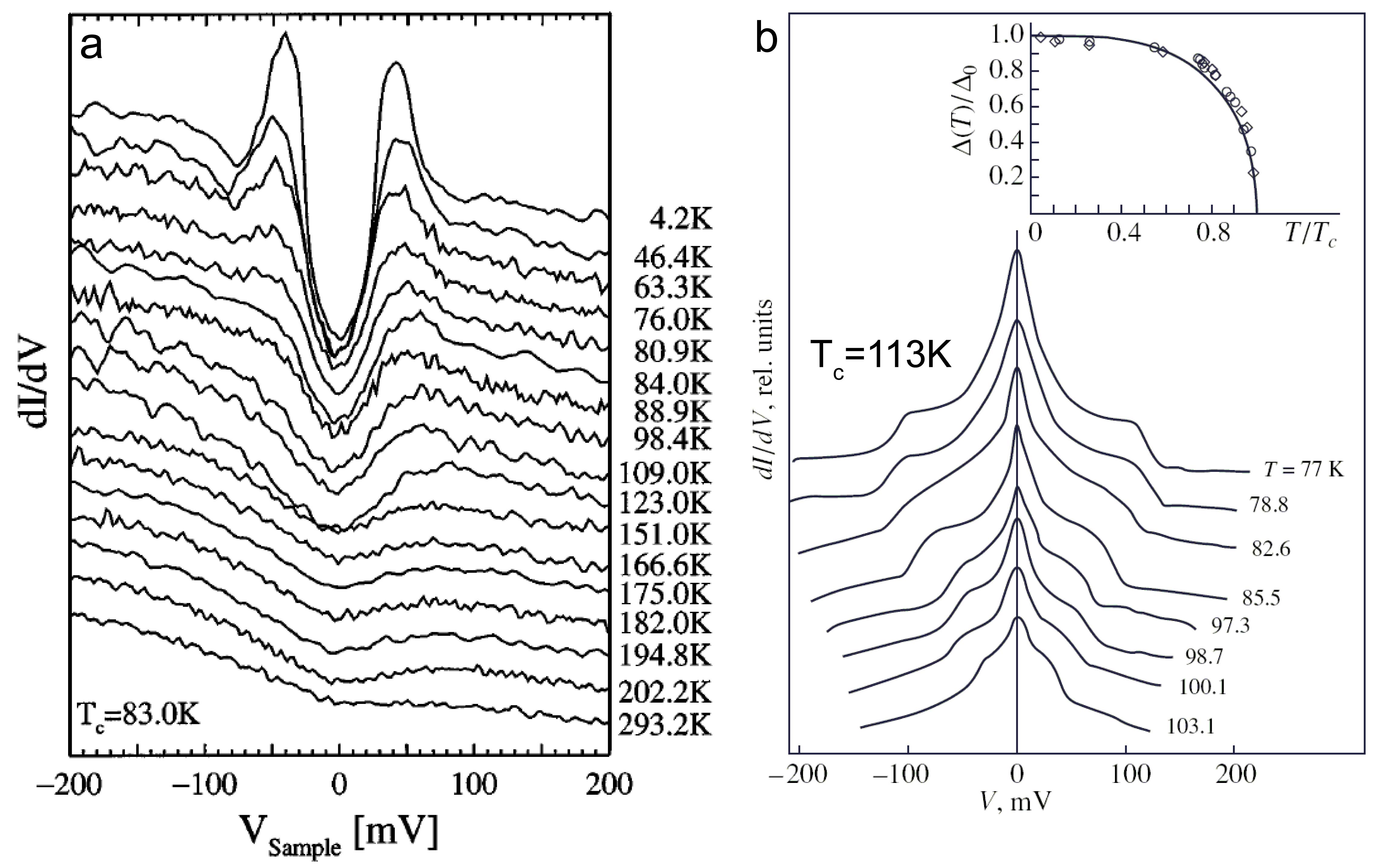}
\centering
\caption{\label{Fig 1: Different temperature dependence}  (a) STS measurements on Bi$_2$Sr$_2$CaCu$_2$O$_{8+\delta}$ (Bi-2212, T$_c$=83) from Ref. \cite{Renner:STM_temp_indept_gap}. (b) Andreev reflection measurements on Bi$_2$Sr$_2$Ca$_2$Cu$_3$O$_{10+\delta}$ (Bi-2223, T$_c$=113) from Ref. \cite{Svistunov:Andreev2223}.  The two experiments imply different temperature dependencies of the gap. }
\end{figure}

The pseudogap and the nodal and antinodal kinks are among the most obvious spectral features in ARPES, directly connected to the two approaches taken to the superconductivity problem, but their origins are a matter of ongoing debate.  Whereas the normal state (T$>$T$_c$) of most low-T$_c$ superconductors can be characterized by Fermi liquid theory, the "normal" state of underdoped cuprates--the pseudogap regime, appearing in ARPES as a partial gapping of the Fermi surface above T$_c$--remains controversial, as different experiments favor conflicting explanations. As an example, we show the incongruous temperature-dependencies of the gap measured by scanning tunneling spectroscopy (STS) and Andreev reflection in Fig. \ref{Fig 1: Different temperature dependence}.  In STS, the gap is defined by the peak-to-peak separation in the tunneling conductance below T$_c$, or by a depletion of density of states (DOS) near zero bias in the normal state.  The STS data in Fig. \ref{Fig 1: Different temperature dependence}(a) suggests that the gap is largely unchanged across T$_c$, and that it persists to temperatures much higher than T$_c$.  Such a temperature dependence is often explained by a 'precursor-pairing' or 'one-gap' scenario where the pseudogap state is attributed to disordered superconductivity: the onset of the pseudogap (T*) marks the onset of Cooper-pair formation, whereas T$_c$ marks the onset of phase coherence.\cite{Kugler:STS_preformedPairs,EmeryKivelson:PhaseFluct,FranzMillis:PhaseFluct}  This scenario is also supported by Nernst effect measurements.\cite{Wang:Nernst} In contrast, Andreev reflection experiments, reproduced in Fig. \ref{Fig 1: Different temperature dependence}(b), indicate a gap with a BCS-like temperature dependence which vanishes near T$_c$.\cite{Svistunov:Andreev2223}  Support for this picture in which the pseudogap state is distinct from superconductivity also comes from Raman scattering.\cite{Raman_Review}  The incompatible temperature dependencies shown in Fig. \ref{Fig 1: Different temperature dependence} serve as a motivation for our experiments on the relation between superconductivity and the pseudogap, and we will show that with the momentum resolution of ARPES, the discrepancy can be resolved.  Meanwhile, it remains a question whether the bosons giving rise to the ubiquitous kinks in ARPES spectra are related to pairing, but before this can be addressed the identity of these bosons--do they have magnetic origin, lattice origin, or something else?--needs to be ascertained.  Complicating this issue are recent observations of many-body interactions at multiple energies, potentially with different origins.\cite{Bi2212NewCouplingLaser:Zhang,HierarchyManyBodyInteractions:Meevasana,Vishik:LEKink} This review is divided into three sections: first, we establish the low-temperature gap and quasiparticle phenomenology in momentum space, and in the second section we review two temperature-dependence studies which demonstrate the distinction between pseudogap and superconductivity phenomena; finally, we discuss low-energy renormalization effects of nodal quasiparticles.

\section{Pseudogap effects below T$_c$}

\begin{figure} [t]
\includegraphics [type=jpg,ext=.jpg,read=.jpg,clip, width=2.8 in]{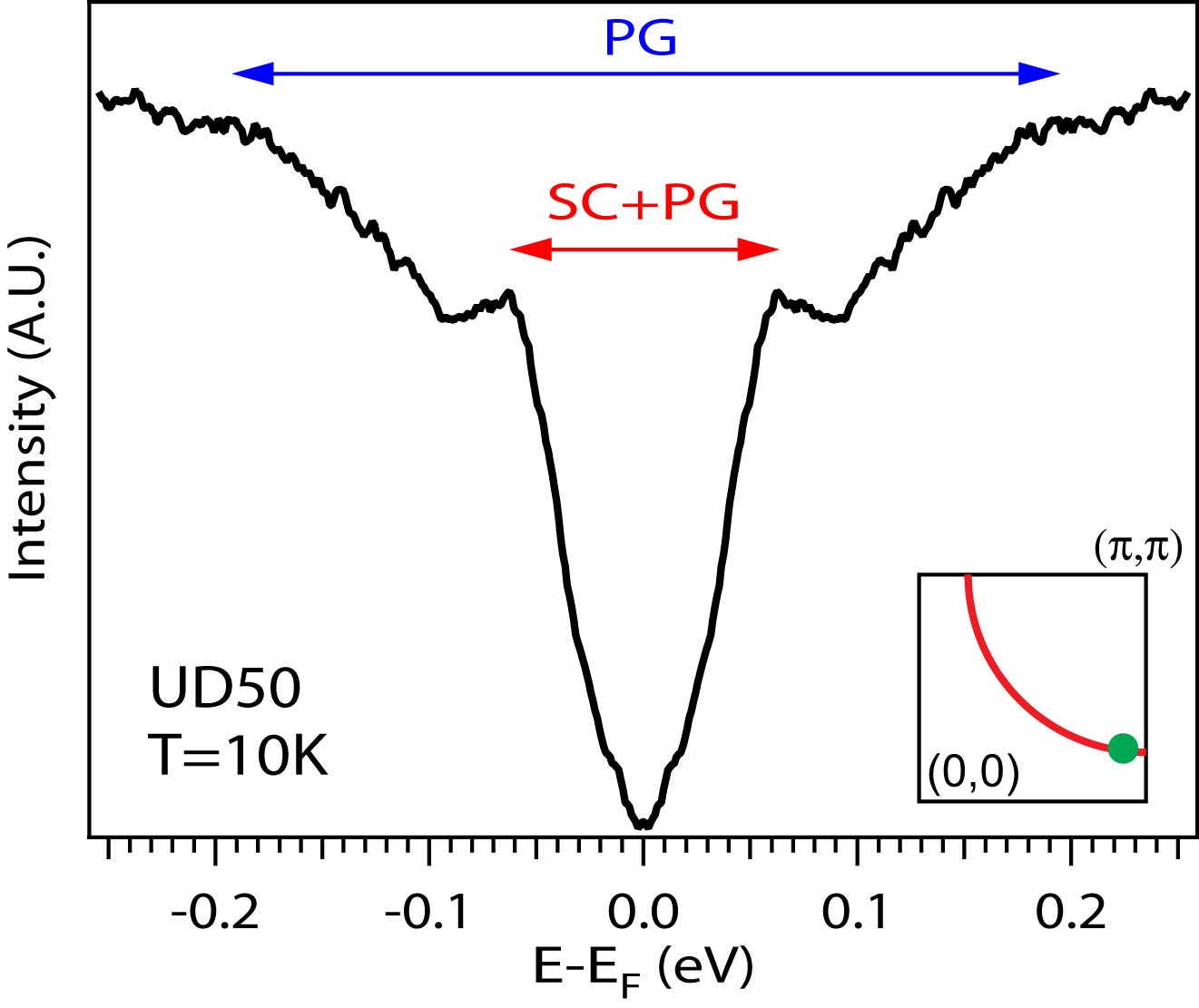}
\centering
\caption{\label{Fig 2: Two Features in EDC}  Symmetrized EDC near the antinode for underdoped Bi-2212 with a T$_c$ of 50K.  Two features are seen in the spectrum: a low-energy peak associated with superconductivity and a broader feature at higher energy associated with the pseudogap.  For such a deeply underdoped system, the intensity and energy position of the superconducting feature is strongly influenced by the underlying pseudogap.}
\end{figure}

Temperatures below T$_c$ mark the realm of superconductivity, but in sufficiently underdoped cuprates, two distinct gaps can be distinguished below T$_c$, both in raw data and in the momentum-dependence of the gap.  Fig. \ref{Fig 2: Two Features in EDC} shows the symmetrized energy distribution curve (EDC, intensity as a function of energy at fixed momentum) near the antinodal k$_F$ for underdoped Bi$_2$Sr$_2$Ca$_{1-x}$Y$_x$Cu$_2$O$_{8+\delta}$  (Bi-2212) with T$_c$$=$50K (UD50, p=8.4$\%$), measured at 10K.  A sharp peak at lower binding energy is associated with superconductivity, and the broader feature at higher binding energy can be attributed to the pseudogap.  We emphasize that the superconducting feature is strongly influenced by the pseudogap, which pushes it to higher binding energy and decreases its intensity.  It should be noted that the peak-dip-hump EDC lineshape seen near ($\pi$,0) in multilayer cuprates is often associated with electron-boson coupling,\cite{ModeCouple:Resonance,Cuk:ReviewEPh_coupling} so it not obvious a-priori why we should characterize the broad hump in Fig. \ref{Fig 2: Two Features in EDC} with the pseudogap.  In principle multiple components--superconductivity, pseudogap, mode-coupling and maybe others--contribute to the antinodal lineshape.  The characterization made in Fig. \ref{Fig 2: Two Features in EDC} is reasonable because of the increasing influence of the pseudogap in the underdoped regime, the doping dependence of the broadly peaked feature at higher energy,\cite{Campuzano:PDH} and the proximity of this larger energy scale to the pseudogap energy scale above T$_c$. This picture will become clear upon studying more experiments, such as the momentum dependence of the superconducting gap.

\begin{figure} [t]
\includegraphics [type=jpg,ext=.jpg,read=.jpg,clip, width=2.5 in]{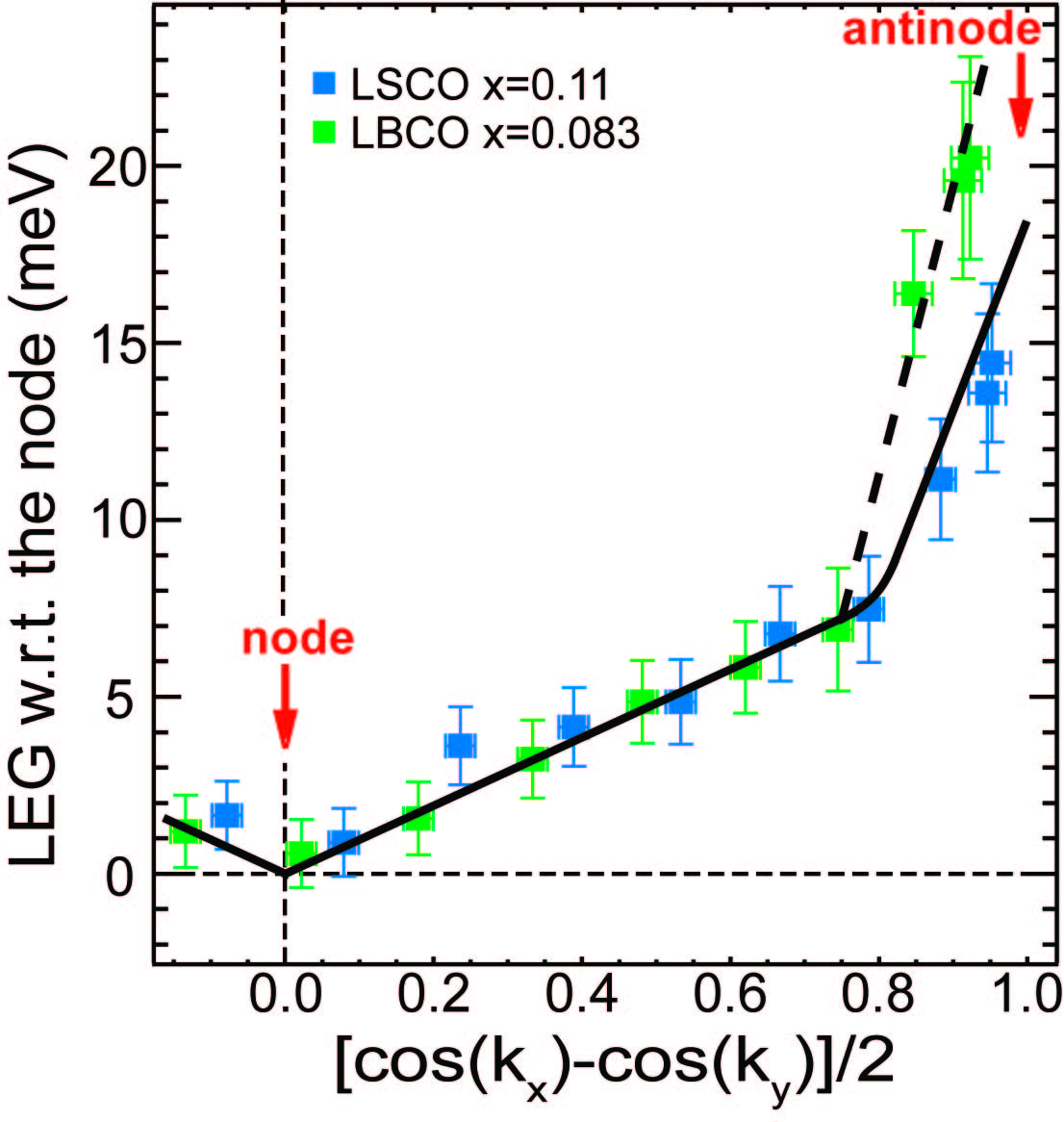}
\centering
\caption{\label{Fig 3: d-wave at low energy}  Leading edge gap function for LBCO x=0.11 (T$_c$=23K) and LSCO x=0.11 (T$_c$=26K) measured at 19$\pm$2K and 21$\pm$2K, respectively, plotted as a function of the simple \textit{d}-wave form, from Ref. \cite{He:LBCO_FailedSC}.  As in other underdoped cuprates, the gap function has a simple \textit{d}-wave form near the node, and deviates from this behavior at the antinode.  This deviation increases as hole concentration is reduced.}
\end{figure}

While a \textit{d}-wave superconducting gap is a hallmark of cuprate high-temperature superconductivity, recent experiments have shown that the gap functions of underdoped systems deviate from a simple \textit{d}-wave form, $\Delta$(\textbf{k})=$|$$\cos$(k$_x$)-$\cos$(k$_y$)$|$/2 near the antinode. \cite{Lee:twoGapARPES_TDep,Tanaka:twoGapARPES_dopingDep,Yoshida:LSCOGapFunction2009,He:LBCO_FailedSC,Kaminski:2gapBi2201}  This is exemplified in Fig. \ref{Fig 3: d-wave at low energy}, which shows the low-temperature gap function of two lanthanum-based cuprates: La$_{2-x}$Ba$_{x}$CuO$_4$ (LBCO) x=0.083 and La$_{2-x}$Sr$_x$CuO$_4$ (LSCO) x=0.11, adapted from Ref. \cite{He:LBCO_FailedSC}.  The gap is extracted from the leading-edge midpoint, a model-independent measure.  Close to the node (the momentum position where the superconducting gap is zero), the gap follows a simple \textit{d}-wave form, but in the antinodal region (momenta near the Brillouin zone axis), the gap increases more rapidly, deviating strongly from the near-nodal momentum dependence.  This deviation is more pronounced for the more underdoped sample.  Thus, the Fermi surface can be divided into two general regions with distinct momentum dependencies of the gap, though we note that the crossover may not be abrupt.

\begin{figure} [t]
\includegraphics [type=jpg,ext=.jpg,read=.jpg,clip, width=2.7 in]{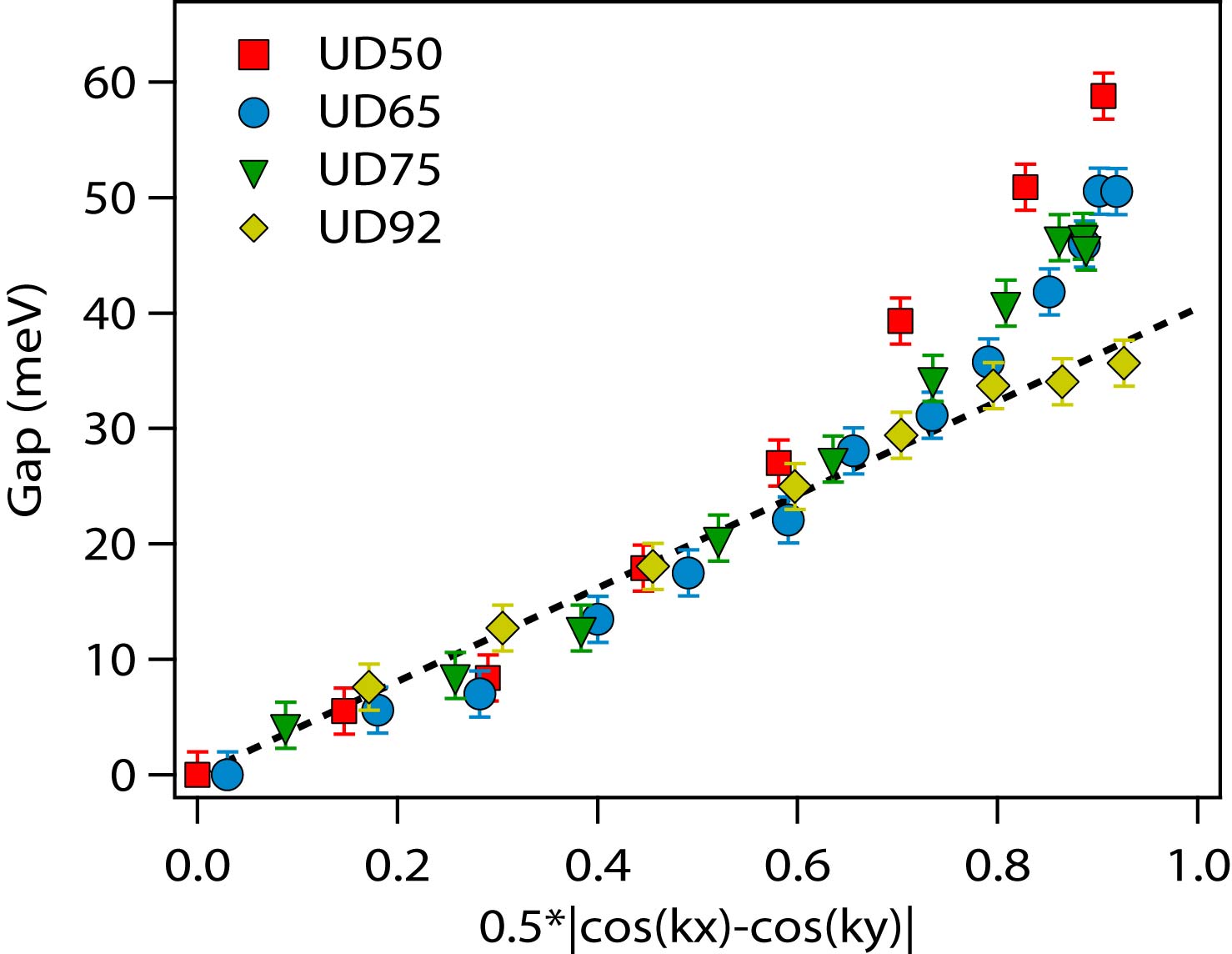}
\centering
\caption{\label{Fig 4: Low temperature gap function in Bi-2212}  Gap function for underdoped Bi-2212, measured at 10K.\cite{Tanaka:twoGapARPES_dopingDep,Lee:twoGapARPES_TDep}  Underdoped samples with T$_c$$<$92K show a deviation from a simple \textit{d}-wave form (dashed line) near the antinode.}
\end{figure}

The same phenomenon is observed in underdoped Bi-2212.  Fig. \ref{Fig 4: Low temperature gap function in Bi-2212} shows the gap functions at 10K for four dopings, partially adapted from Refs. \cite{Lee:twoGapARPES_TDep} and \cite{Tanaka:twoGapARPES_dopingDep}.  At each momentum, gaps were extracted by fitting symmetrized EDCs to a minimum model.\cite{Symmetrization_Norman_model}.  The sample with the largest hole concentration in this Figure (UD92K) follows a simple \textit{d}-wave form all around the Fermi surface at low temperature.  As with lanthanum-based cuprates, more underdoped systems show increasing deviation from this form in the antinodal region, with the most underdoped sample (UD50) showing a deviation of more than 20 meV.  It is well established that the pseudogap energy scale increases with underdoping,\cite{PhaseDiagramInPlaneResistivity:Ando,DopingDependencePG:Tallon,Campuzano:PDH} so it is natural to associate the increasing deviation from a simple \textit{d}-wave gap form with the increasing influence of pseudogap physics in the superconducting state of underdoped cuprates.  Notably, the near-nodal region lacks this strong doping dependence in this doping regime.

\begin{figure*} [t]
\includegraphics [type=jpg,ext=.jpg,read=.jpg,clip, width=5.5 in]{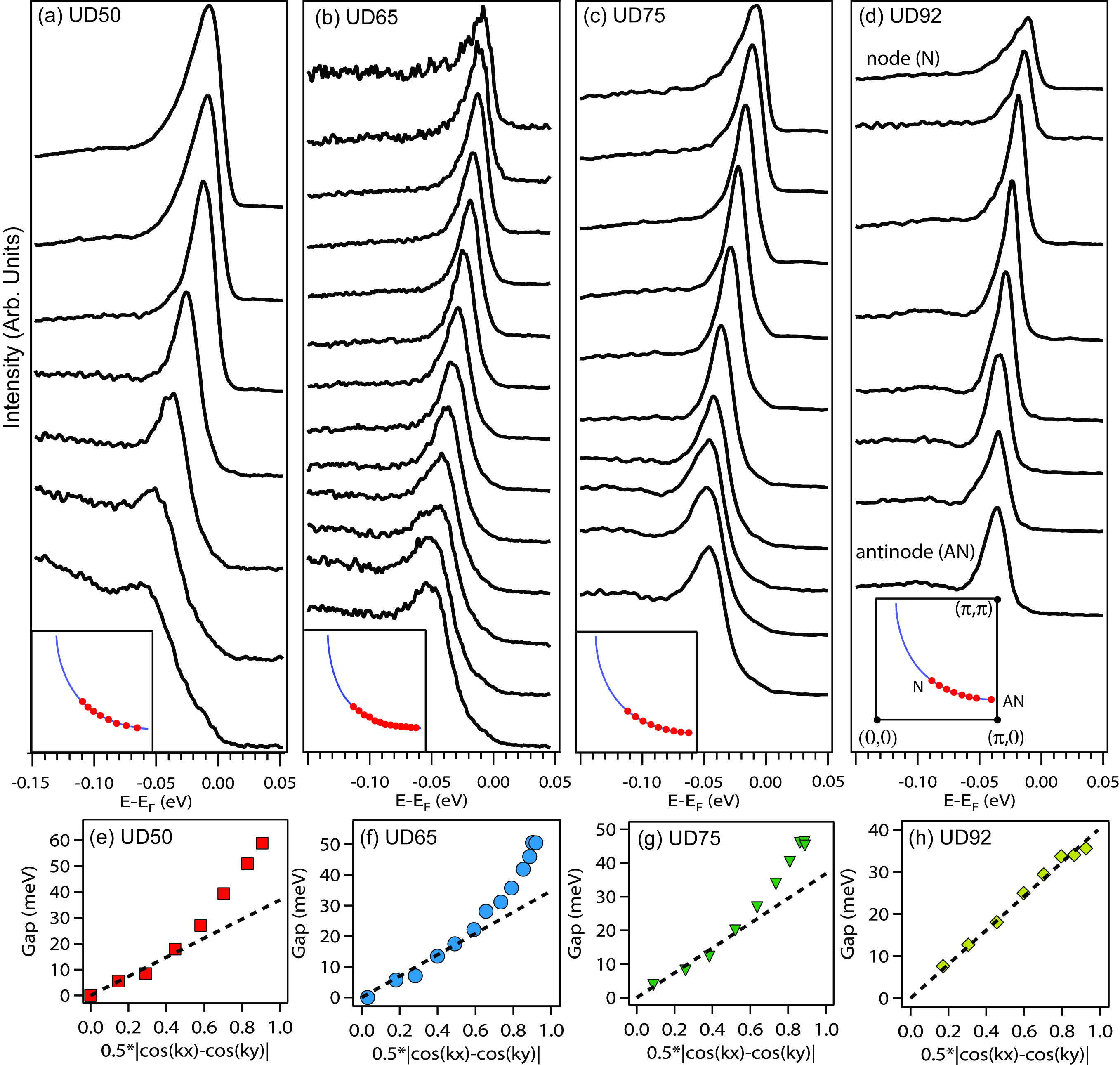}
\centering
\caption{\label{Fig 5: Low-temperature EDCs}  (a)-(d) EDCs at k$_F$ (T=10K) for four dopings.  Top curve is near the node and bottom curve is near the antinode.  Insets sketch where cuts intersect Fermi surface.  Sharp peak in all EDCs demonstrate that quasiparticles are observed by ARPES all around the Fermi surface in underdoped Bi-2212 in this doping regime.  (e)-(h) Gaps fit from the EDCs in (a)-(d).  The three most underdoped data show a deviation from a simple \textit{d}-wave form near the antinode, while still having sharp quasiparticle peaks at the same momenta.}
\end{figure*}

Some have argued that a deviation from a simple \textit{d}-wave form is an artifact from the loss of sharp quasiparticles in the antinodal region.\cite{Wei:CoherencePeak2201}  However, in Bi-2212, quasiparticles are ubiquitous all around the Fermi surface, well into the underdoped regime (e.g. UD50), so the criteria for extracting the gap is identical in the near-nodal and near-antinodal regions.  In Fig. \ref{Fig 5: Low-temperature EDCs}(a)-(d) we plot the EDCs at k$_F$ for different momenta between the node and the antinode for four different underdoped Bi-2212 samples.  Quasiparticle peaks--sharp peaks at low binding energy-- are visible all around the Fermi surface, and most notably in the antinodal region, for all the dopings in the figure.  Fig. \ref{Fig 5: Low-temperature EDCs}(e)-(h) show the corresponding gap functions.  For the three most underdoped samples, the gap near the antinode deviates from a simple \textit{d}-wave form, yet quasiparticle peaks still persist in the momentum region of this deviation. As doping decreases, the peak near the antinode becomes weaker relative to the features at higher binding energy, as previously reported by Feng \textit{et al},\cite{Feng:QPScience} suggesting that superconductivity becomes weaker as the antinodal pseudogap becomes stronger, hinting at competition between two distinct states.  Though antinodal quasiparticle peaks have not been observed for all cuprates, LSCO (p=0.15) provides another example of a system where a deviation from a simple \textit{d}-wave form is accompanied by quasiparticle-like peaks all around the Fermi surface.\cite{Yoshida:LSCOGapFunction2009}

STS measures the local DOS averaged over all momenta, so tunneling spectra can be interpreted as a momentum-space average of ARPES data, neglecting possible matrix element effects.  The DOS at low bias voltage corresponds to the near-nodal region, as the antinodal region is completely gapped at low energy, and the peaks in the spectra (see Fig. \ref{Fig 1: Different temperature dependence}(a)) are more influenced by antinodal states.  Notably, two features are seen in STS spectra as well.  Recent experiments have shown a hump in the DOS at low bias voltage ($<$50meV), in addition to the higher energy peaks from from which the gap is extracted (see for example Fig. 1 of Ref. \cite{Pushp:UniversalNode} or Fig. 2(a) of Ref. \cite{McElroy:ANDecoherenceCheckerboard}).  These two features correspond well with the two 'features' seen in momentum space by ARPES: a simple \textit{d}-wave gap near the node, and a larger gap near the antinode.

\begin{figure} [t]
\includegraphics [type=jpg,ext=.jpg,read=.jpg,clip, width=5 in]{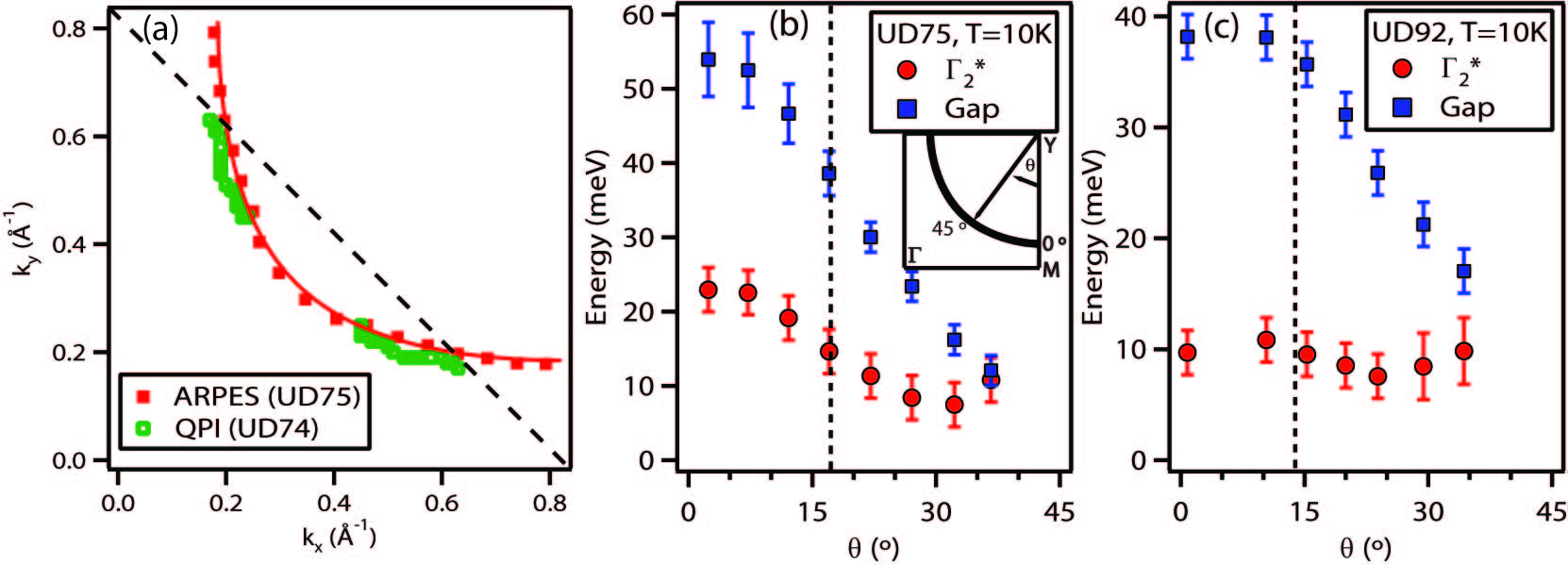}
\centering
\caption{\label{Fig 6: Comparison of ARPES and QPI}  (a) Locus of quasiparticles observed by ARPES and QPI (Refs. \cite{Kohsaka:extinction,Vishik:QPI_ARPES}) for samples of similar T$_c$.  ARPES observes sharp quasiparticles all around the FS, whereas QPI implies quasiparticle termination at the antiferromagnetic zone boundary (dashed line). (b)-(c) Gap and scattering rate around the Fermi surface for UD92 and UD75, demonstrating that all the peaks in Fig. \ref{Fig 5: Low-temperature EDCs}(c)-(d) are quasiparticle-like with a smoothly-evolving scattering rate.  Details of the fitting described in Ref. \cite{Vishik:QPI_ARPES}}
\end{figure}

STS has increasingly exploited the phenomenon of quasiparticle interference (QPI) to learn about momentum-space properties of cuprates.\cite{Kapitulnik:DOS_modulationSTM,Kohsaka:extinction,McElroy:exctinction,Wise:QPI_localInhomogeneity}  Quasiparticles scattering from impurities in a superconductor interfere with one another, producing a standing wave pattern in the local density of states $\rho$(\textbf{r},$\omega$), which can be studied via Fourier transform, $\rho$(\textbf{q},$\omega$).  The dispersion of the peaks in $\rho$(\textbf{q},$\omega$) as a function of bias voltage $\omega$ is analyzed in terms of the octet model and yields information about the Fermi surface and momentum dependence of the superconducting gap.\cite{Kohsaka:extinction,McElroy:exctinction,Wise:QPI_localInhomogeneity}  Recent FT-STS studies reported that at low bias voltage, the dispersion of $\rho$(\textbf{q},$\omega$) behaves as expected for a \textit{d}-wave superconductor, but upon reaching the bias voltage associated with the antiferromagnetic (AF) zone boundary (line connecting ($\pi$,0) and (0,$\pi$)), many of the peaks in $\rho$(\textbf{q},$\omega$) disappear, leaving behind a localized state which breaks translational and rotational symmetry, which has been associate with pseudogap physics.\cite{Kohsaka:extinction}  From the extinction of QPI at the AF zone boundary, it has been suggested that superconducting quasiparticles themselves become extinct at the AF zone boundary, even in overdoped materials which are far away from the parent AF Mott-insulator state on the phase diagram.  Fig. \ref{Fig 5: Low-temperature EDCs}(a)-(d) clearly refutes the claim of extinction, as sharp peaks are seen all the way to the antinode, and Fig. \ref{Fig 6: Comparison of ARPES and QPI}(b)-(c) show that these peaks are always quasiparticle-like (scattering rate is smaller than binding energy) and their fitted scattering rate evolves smoothly around the Fermi surface, without diverging near the AF zone boundary.

On one hand, the local DOS measured by STS appears very consistent with the momentum dependence of the gap in Bi-2212, but on the other hand, the Fourier transform yields inconsistencies regarding the momentum-dependence of the quasiparticles.  As discussed in Ref. \cite{Vishik:QPI_ARPES}, this inconsistency can be resolved by considering the different natures of the probes: ARPES studies the single-particle spectral function, while Fourier-transform STS studies the quasiparticle interference pattern, which represents a two particle process.  Thus, quasiparticles can exist all around the Fermi surface but not contribute to the interference pattern in certain regions.  It should be noted that the work of Kohsaka \textit{et al.} captured some key features of underdoped cuprates:  the pseudogap state is present even below T$_c$ and it resides in the antinodal region of momentum space, with superconductivity dominating closer to the node.  However, the ubiquity of quasiparticles in Bi-2212 (Fig. \ref{Fig 5: Low-temperature EDCs}) suggests that the two orders have a complex interplay in momentum space.  In the next section we will discuss two temperature dependence studies which further distinguish the pseudogap from superconductivity and offer clues to the origin of the pseudogap state.

\section{Temperature Dependence}
\begin{figure} [t]
\includegraphics [type=jpg,ext=.jpg,read=.jpg,clip, width=5 in]{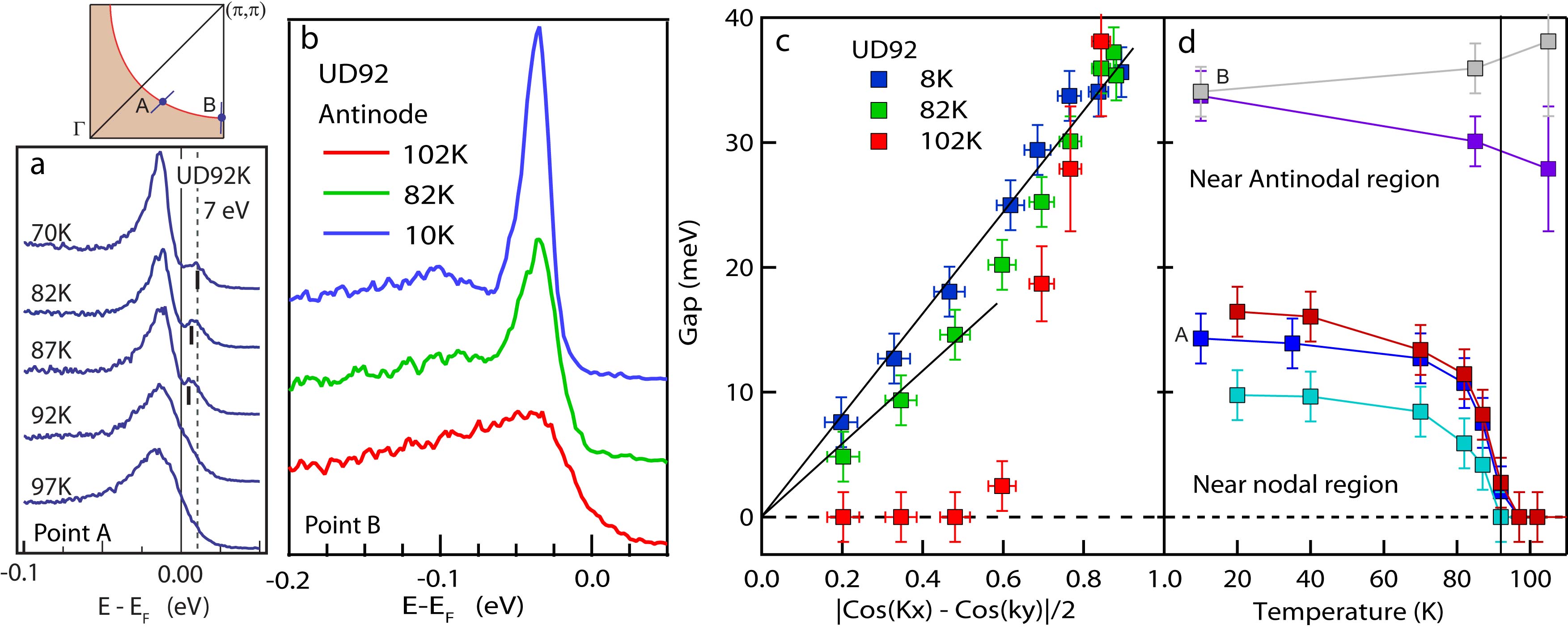}
\centering
\caption{\label{Fig 7: Temperature dependence of gap in Bi-2212} (a) From Ref. \cite{Lee:twoGapARPES_TDep}. Underdoped Bi-2212, T$_c$$=$92K: EDCs near T$_c$ for point A, marked in the Fermi surface schematic.  At elevated temperature, the upper Bogoliubov branch, a signature of superconductivity, is thermally populated, and its energy position is marked by a short vertical line.  The upper Bogoliubov peak moves closer to E$_F$ as T$_c$ is approached.  (b) EDCs at the antinode (point B) for the three temperatures.  Though the position of the EDC maximum remains unchanged, the sharp peak disappears across T$_c$. (c) Gap function at three different temperatures.  In the near-nodal region, the gap evolves as it approaches T$_c$, leaving an ungapped 'Fermi arc' above T$_c$, whereas the gap in the antinodal region does not evolve substantially across T$_c$ (d) Temperature dependence of the gap in the near-nodal and the near-antinodal regions, illustrating dichotomy.}
\end{figure}

Early ARPES studies on underdoped Bi-2212 found a \textit{d}-wave gap below T$_c$ and a partially-gapped state above T$_c$, where the Fermi surface was restored only in a limited Fermi arc near the node.\cite{DestructionFS_Norman} Recently, Lee \textit{et al} have shown that the details of this temperature evolution yield crucial information about the relation between the pseudogap and superconductivity.\cite{Lee:twoGapARPES_TDep}  Fig. \ref{Fig 7: Temperature dependence of gap in Bi-2212}(a)-(b) shows the temperature dependence of EDCs in the near-nodal and antinodal regions respectively.  When a superconducting gap opens on the Fermi surface, the band dispersion splits into the upper and lower Bogoliubov branches.  At low temperature, only the latter is measured by ARPES because of the sharp Fermi-Dirac cutoff, but at elevated temperatures, a portion of the unoccupied spectrum close to the Fermi level is thermally populated and can be studied by ARPES as well.  In the near-nodal region the upper Bogoliubov branch at k$_F$ appears as a peak above E$_F$.  As the temperature is raised towards T$_c$, the position of this peak moves closer to E$_F$, and it disappears altogether close to T$_c$.  At the antinode, however, the position of the peak below E$_F$ is largely independent of temperature, but there is a profound change in lineshape across T$_c$: a sharp peak is present below T$_c$ (82K), but only a broadly peaked feature is seen above T$_c$ (102K).

The temperature dependence in the raw EDCs is quantified by fitting to a minimal model, as described in Ref. \cite{Lee:twoGapARPES_TDep}.  Fig. \ref{Fig 7: Temperature dependence of gap in Bi-2212}(c) shows the momentum dependence of the gap in Bi-2212 UD92 at three different temperatures. At 10K, the gap function follows a simple \textit{d}-wave form, and at 102K a gap function characteristic of the pseudogap above T$_c$ is measured:  an ungapped arc near the node and a gap at the antinode whose energy is comparable to the antinodal gap below T$_c$.  At intervening temperatures, the gap evolves more rapidly in the near-nodal region, giving rise to a deviation from a simple \textit{d}-wave form at 82K.  Fig. \ref{Fig 7: Temperature dependence of gap in Bi-2212}(d) details the temperature dependence in the near-nodal and near-antinodal region and points out an interesting dichotomy: in the near-nodal region, the gap has a strong temperature dependence, and closes at a temperature near T$_c$, whereas the gap in the antinodal region is unchanged across T$_c$.  Because of this differing temperature dependence, it is unlikely that the near-nodal region and the antinodal region can be associated with a single order parameter.  Thus we posit that the former is dominated by superconductivity and the latter by pseudogap physics.  However, the presence of quasiparticles all around the Fermi surface suggests a subtle interaction between these two states in momentum space.

With these observations, we can re-examine the discrepancy between tunneling and Andreev reflection experiments introduced in Fig. \ref{Fig 1: Different temperature dependence}.  Andreev reflection reported a temperature-dependent gap closing near T$_c$, similar to the gap measured by ARPES in the near-nodal region, whereas STS reported a temperature-independent gap, similar to the ARPES gap in the antinodal region.  From the momentum-resolution of ARPES,  we can reconcile these experiments by recognizing that STS and Andreev reflection are likely sensitive to different portions of the Fermi surface in the cuprates.  Moreover, because Andreev reflection couples directly to the superconducting condensate, the temperature dependence reported in Ref. \cite{Svistunov:Andreev2223} may be the generic temperature dependence of the superconducting gap, and consistent temperature-dependent ARPES experiments support the idea that superconducting order dominates in the near-nodal region.

\begin{figure} [t]
\includegraphics [type=jpg,ext=.jpg,read=.jpg,clip, width=5 in]{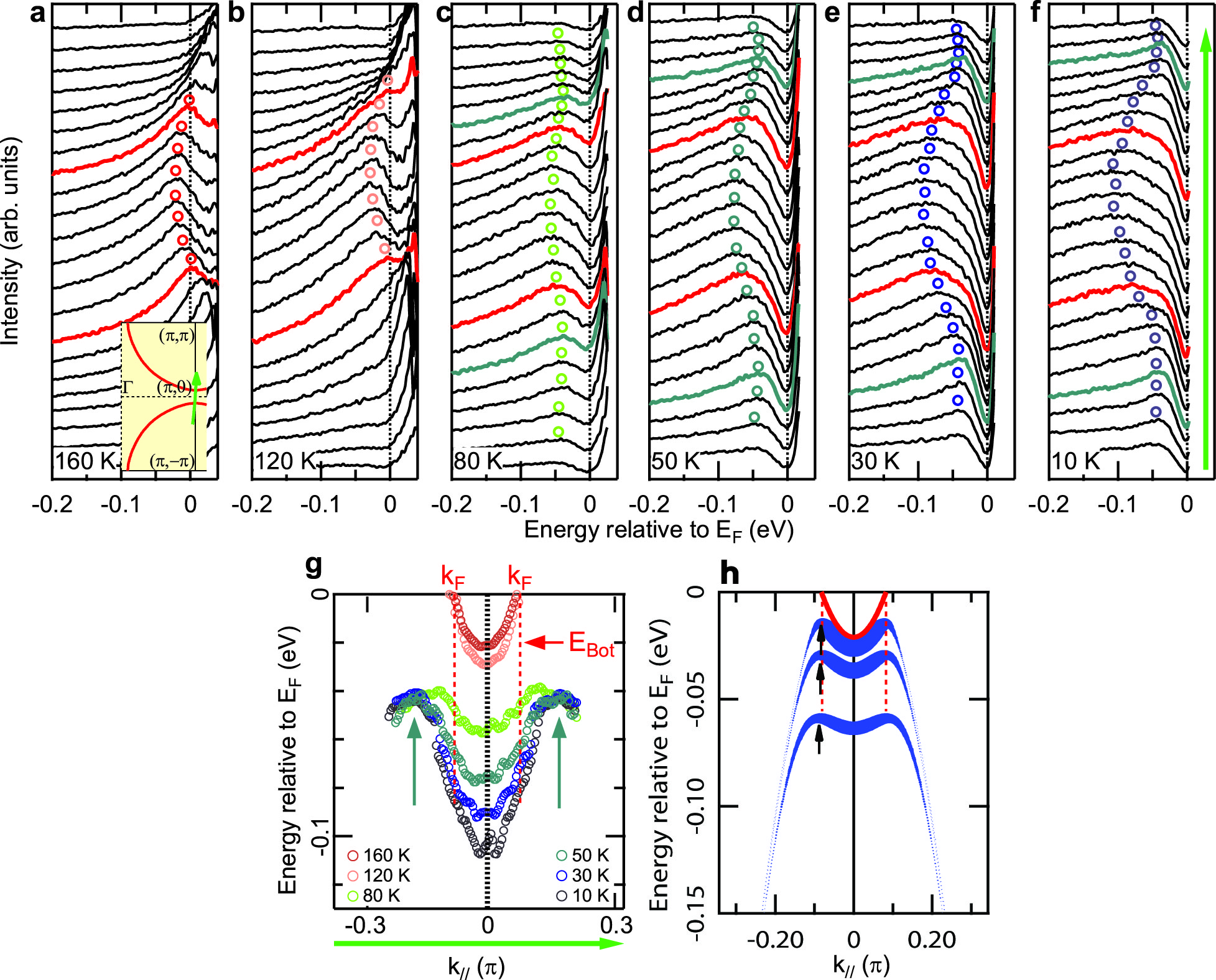}
\centering
\caption{\label{Fig 8: Temperature dependence of EDCs in Bi-2201}  (a)-(f) EDCs in the antinodal region of optimally doped Pb-Bi2201 (T$_c$=34K, T*=125K), along the cut marked by the green arrow in the inset of (a), divided by the Fermi-Dirac function convolved with instrument resolution.  At the highest temperature (160K, a), a simple parabolic band is recovered, and its Fermi crossing points are used to define k$_F$ which is highlighted in red in (a)-(f).  Circles mark the positions of EDC maxima, which are used to define the band position.  At the lowest temperature (f), the momentum position where the band position is closest to E$_F$ (the back-bending position, highlighted in blue) is far removed from k$_F$.  Additionally, the low-temperature spectra are anomalously broad. (g) Summary of the band positions from the EDCs in (a)-(f).  Both the band bottom, and the back-bending positions (marked by vertical arrows) show a temperature evolution. (h) Simulated band position in the presence of superconducting gaps of magnitudes 15, 30, and 60 meV (blue curves, from top to bottom).  Red curve denotes ungapped, parabolic band.  Vertical dashed lines mark k$_F$ and arrows mark back-bending momenta, which are coincident with k$_F$ in the presence of only a superconducting order. }
\end{figure}

In optimally- or under-doped Bi-2201, the quasiparticle peak at the antinode can be weak or absent, \cite{Sato:LayerDep} making it an ideal system to study the underlying pseudogap near the antinode directly.\cite{HeHashimoto:Bi2201_eh_sym_break}  Moreover, it lacks complications from bilayer splitting\cite{Feng:BilayerSplitting} or certain many-body effects,\cite{Lee:LayerEffectsThallium} which may obfuscate the antinodal spectra of multilayer cuprates.  Fig. \ref{Fig 8: Temperature dependence of EDCs in Bi-2201} (a)-(f) show EDCs along an antinodal cut in optimally doped Pb$_{0.55}$Bi$_{1.5}$Sr$_{1.6}$La$_{0.4}$CuO$_{6+\delta}$ (Pb-Bi-2201, T$_c$=34K, T*=125$\pm$10K).   Remarkably, the EDCs at 160K are much sharper than those at 10K, opposite to conventional thermal broadening.  This suggests that something must intervene at intermediate temperatures to produce such broad features at low temperature.  The momentum-dependence of the EDC maxima define the band dispersion near ($\pi$,0).  At the highest temperature (160K), the band is parabolic and has well-defined Fermi crossings, which define k$_F$.  At the three lowest temperatures, the spectra are gapped at all momenta in the cut, and the position where the band comes closest to E$_F$ (the back bending position) differs markedly from k$_F$ established at higher temperature.  The summary of the EDC maxima at different temperatures is plotted in Fig. \ref{Fig 8: Temperature dependence of EDCs in Bi-2201}(g), showing the temperature evolution of the ARPES derived band structure.  At low temperature, the band bottom is at higher binding energy than at high temperature, suggesting that a gap opens between 160K and 80K, but the mismatch between the high-temperature k$_F$ and the low-temperature back-bending position is not consistent with the opening of a superconducting gap.  This mismatch is evidence that the low-temperature antinodal state breaks particle-hole symmetry.

\begin{figure} [t]
\includegraphics [type=jpg,ext=.jpg,read=.jpg,clip, width=3.5 in]{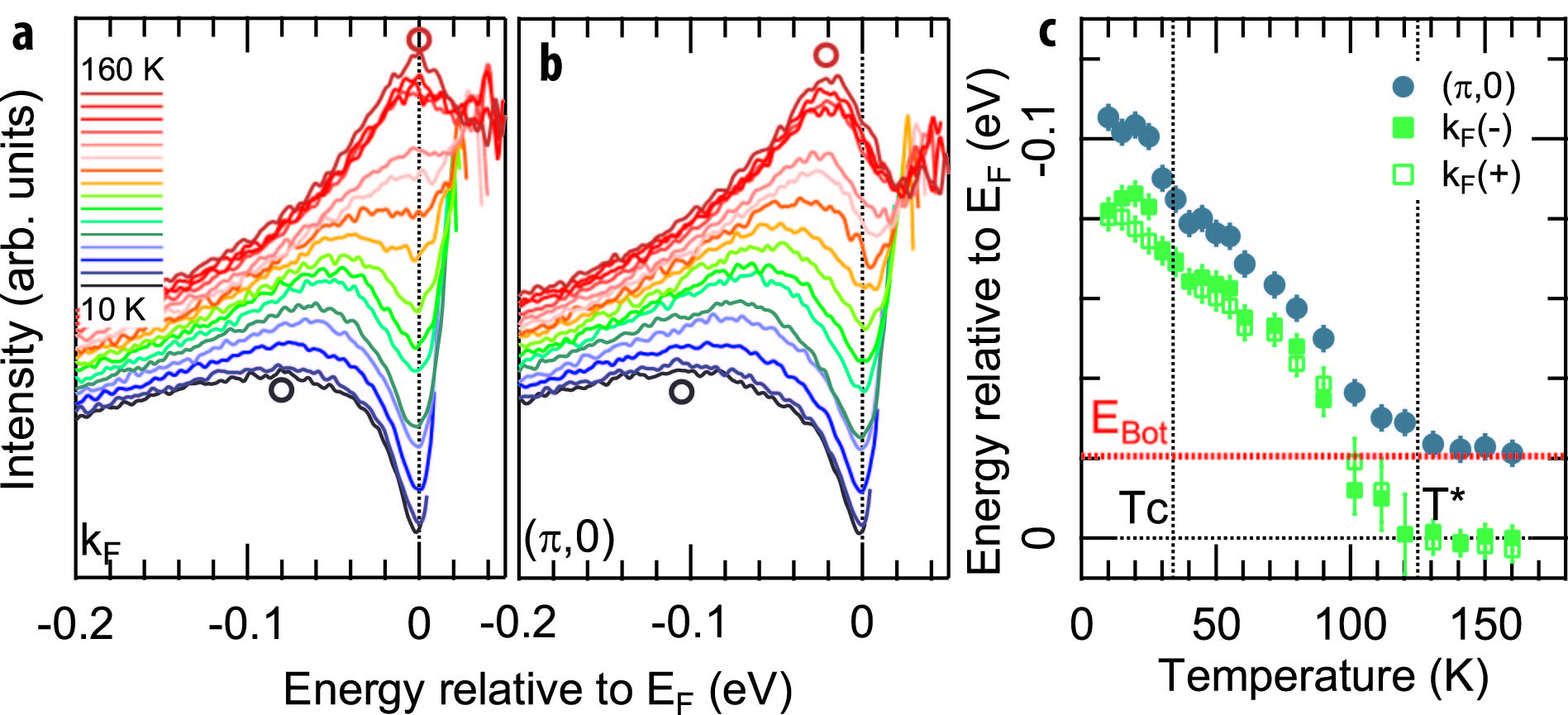}
\centering
\caption{\label{Fig 9: Temperature dependence of Bi-2201 spectra} (a)-(b) Temperature dependence of EDCs at k$_F$ and ($\pi$,0) from 10K (bottom) to 160K (top) in optimally doped Pb-Bi2201.  As temperature is increased, both spectra move to lower binding energy. (c)Temperature dependence of EDC maxima at k$_F$ and ($\pi$,0), at T* the former reaches zero and the latter reaches the band bottom energy.}
\end{figure}

The temperature evolution of the antinodal spectra can constrain the explanation for this phenomenology.  The temperature dependence of the EDCs at k$_F$ and ($\pi$,0) is shown in Fig. \ref{Fig 9: Temperature dependence of Bi-2201 spectra}.  As temperature is raised, the EDC maxima at both momenta move to lower binding energy, and at about 125K, the EDC maximum at k$_F$ reaches E$_F$, marking the closure of the antinodal gap, and confirming T*.  At this temperature, the EDC maximum at ($\pi$,0) also reaches its terminal position, the band bottom.  This temperature-dependent opening of the gap, measured both at k$_F$ and the band bottom, presents a different picture from some previous spectroscopy measurements which reported a temperature-independent antinodal gap,\cite{Symmetrization_Norman_model,Nakayama:GapBi2201_Tdep,Kugler:STS_preformedPairs} and our result suggests a different paradigm for understanding the pseudogap.

\begin{figure} [t]
\includegraphics [type=jpg,ext=.jpg,read=.jpg,clip, width=3.5 in]{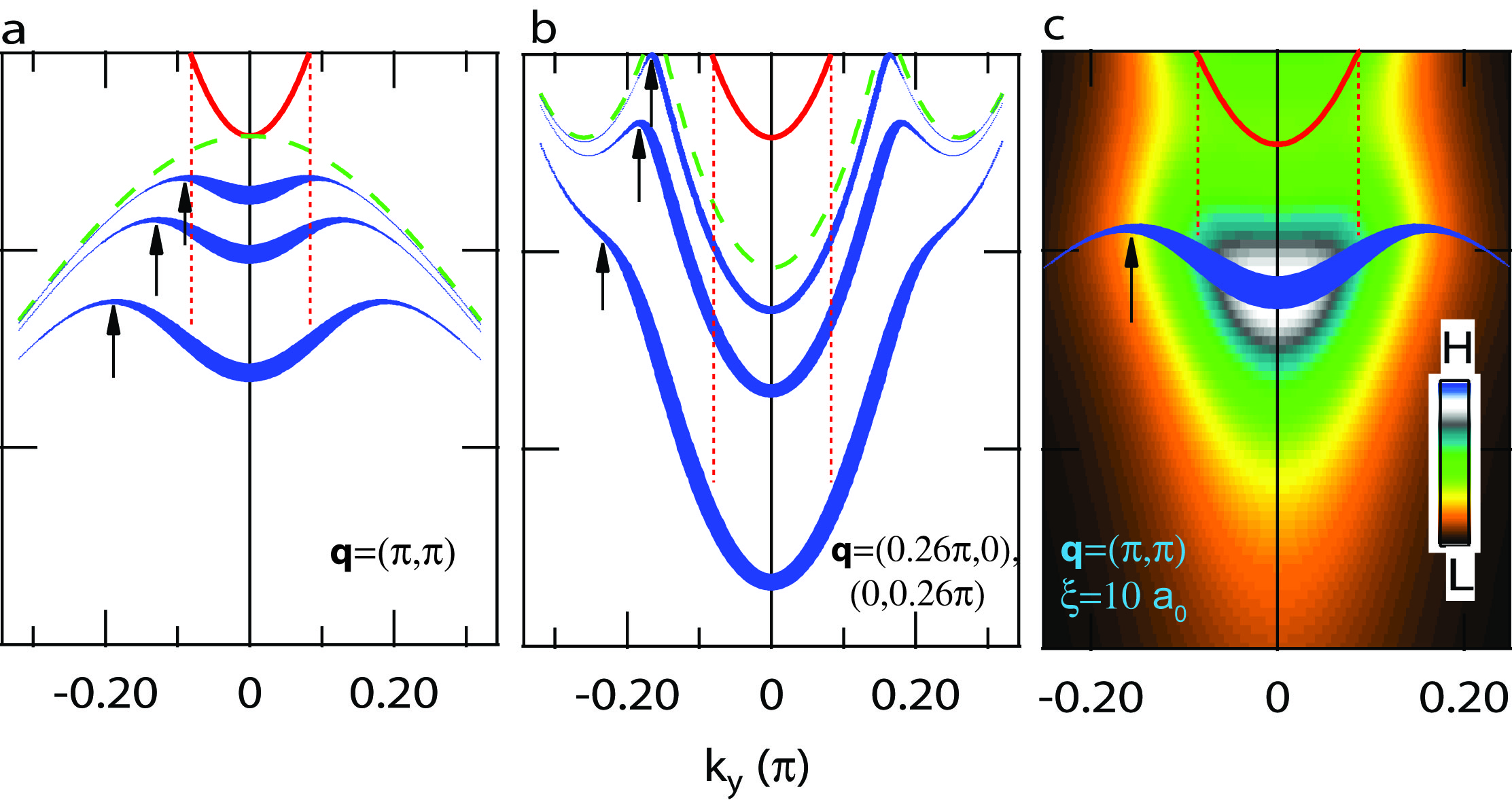}
\centering
\caption{\label{Fig 10: Simulated spectra for example orderings} Simulated spectra along ($\pi$, $\pi$)-($\pi$,0)-($\pi$, -$\pi$) for (a) a commensurate ($\pi$, $\pi$) density wave and (b) an incommensurate checkerboard density wave order with orthogonal wavevectors (0.26$\pi$,0) and (0,0.26$\pi$).  The green dashed curves are shadow bands due to each order which interact with the bare band (red) to produce the renormalized bands (blue).  Three renormalized bands are plotted, corresponding to gaps of 15, 30, and 60meV respectively.  Red vertical dashed lines correspond to k$_F$ and vertical arrows correspond to back bending position. (c) a ($\pi$, $\pi$) density-wave order with finite correlation length $\xi$=10 lattice constants (a$_0$), and density-wave gap $\Delta$$_k$ $=$ 60 meV, independent of \textit{k}.  The fluctuating density wave order qualitatively captures the anomalous spectral broadening seen experimentally.}
\end{figure}

What could explain such unusual low-temperature spectra, which are anomalously broad and exhibit a mismatch between k$_F$ and the back-bending momentum?  We have already shown that the latter feature cannot be captured by superconductivity alone (Fig. \ref{Fig 8: Temperature dependence of EDCs in Bi-2201}(h)).  On the other hand, when a gap opens as a consequence of band folding from a density wave order, the back-bending position can differ from k$_F$, as shown in Fig. \ref{Fig 10: Simulated spectra for example orderings}(a)-(b) for two different ordering wavevectors.  However, a long-range density wave order will not reproduce the low-temperature spectral broadening. This aspect can be captured by considering a fluctuating order, as shown in Fig. \ref{Fig 10: Simulated spectra for example orderings}(c).  Such a local density wave picture is supported by STS experiments, which have reported a symmetry breaking state at pseudogap energies.\cite{Kohsaka:extinction}

\begin{figure} [t]
\includegraphics [type=jpg,ext=.jpg,read=.jpg,clip, width=3.8 in]{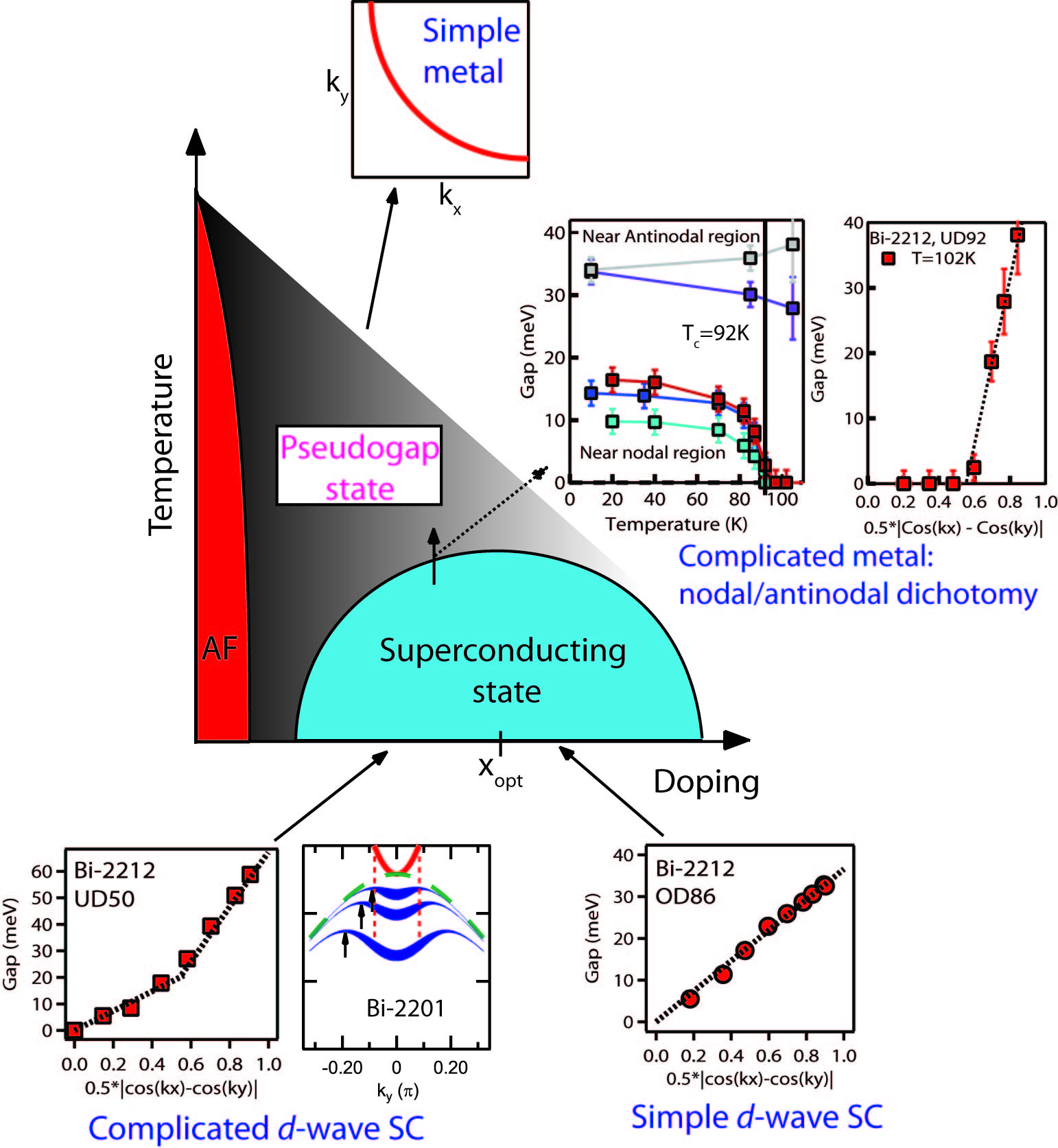}
\centering
\caption{\label{Fig 11: Cartoon summary of temperature and doping dependence} Summary of temperature- and doping-dependence ARPES studies.  In the overdoped regime and above T*, behavior is conventional, with a simple \textit{d}-wave form of the superconducting gap in the former and a simple metallic Fermi surface in the latter.  In the underdoped regime, the low-temperature gap function deviates from a simple \textit{d}-wave form, showing strong influence of the underlying pseudogap on the energy positions of the near-antinodal quasiparticle peaks.  In Bi-2201, the pseudogap state was found to be compatible with a density wave order.  Experiments on slightly underdoped Bi-2212 have found that the antinodal gap shows minimal temperature dependence across T$_c$, but in the near-nodal region the gap collapses near T$_c$, leading to the mysterious pseudogap state where the near-nodal region is metallic and the near-antinodal region is gapped.}
\end{figure}

Our ARPES studies on the superconducting gap and pseudogap in the cuprates reveal an interesting and subtle phenomenology, which is summarized in Fig. \ref{Fig 11: Cartoon summary of temperature and doping dependence}.   We argue that superconductivity and the pseudogap constitute distinct orders based on their coexistence below T$_c$ and the different temperature dependence of the gap at different momenta.  Bi-2201 provides an example of a system where the anomalous behavior of the pseudogap is consistent with a density wave order.

\section{Low-energy nodal excitations}
In this section, we turn to the subject of electron-boson coupling, focusing along the nodal direction [(0,0) to ($\pi$, $\pi$)], where the kink can be studied without complications from gapped spectra and strong bilayer splitting.  The most famous of these features is the ubiquitous dispersion kink seen at 50-80 meV in all cuprates, characterized by a sudden change in velocity and accompanying change in linewidth.\cite{ElectronPhononCoupling:Lanzara,KinkPRL:Bogdanov,Zhou:EPhCoupling,Cuk:ReviewEPh_coupling,Johnston:ReviewRenormalization} Despite the ubiquity of the kink, its explanation remains controversial, with the leading theories being of magnetic \cite{Kaminski:kinkResonanceMode,Borisenko:ResonanceMode} or phonon origin.\cite{MultipleModes:XJ, Devereaux:AnisotropicEph}  There have been proposals that the bosons associated with this kink may be connected to superconductivity, either by mediating pairing or by enhancing T$_c$,\cite{Dahm:SpinFluctPairing,Johnston:SystematicStudyEPhCoupling} but others have questioned whether cuprate superconductivity needs a pairing glue.\cite{Anderson:NoGlue}  Our experiments do not directly address the issue of whether the nodal kinks arise from the same phenomena which bind Cooper pairs, but they propose an origin for the ubiquitous 50-80 meV kink, and demonstrate the ubiquity and doping dependence of another kink at lower energy.  By recognizing and understanding the multitude of many body phenomena in the cuprates, we can make progress towards solving the superconductivity puzzle.

In this section, we will focus on the kinks in Bi-2212, the most prominent of which appears at 70 meV.   For a \textit{d}-wave superconductor with a maximum gap $\Delta$$_0$, coupling to a mode of energy $\Omega$ is expected to produce a nodal dispersion kink at an energy $\Omega$$+$$\Delta$$_0$.\cite{ElectronPhonon:Sandvik}  Near optimal doping, $\Delta$$_0$ $\sim$ 35-40 meV, so in the superconducting state, the nodal kink is often attributed to a bosonic mode with an energy near 40 meV, such as the B$_{1g}$ bond buckling phonon \cite{Cuk:B1gAN} or the magnetic resonance mode,\cite{Gromko:Eboson_couple} gap-shifted by $\Delta$$_0$.

\begin{figure} [t]
\includegraphics [type=jpg,ext=.jpg,read=.jpg,clip, width=4.6 in]{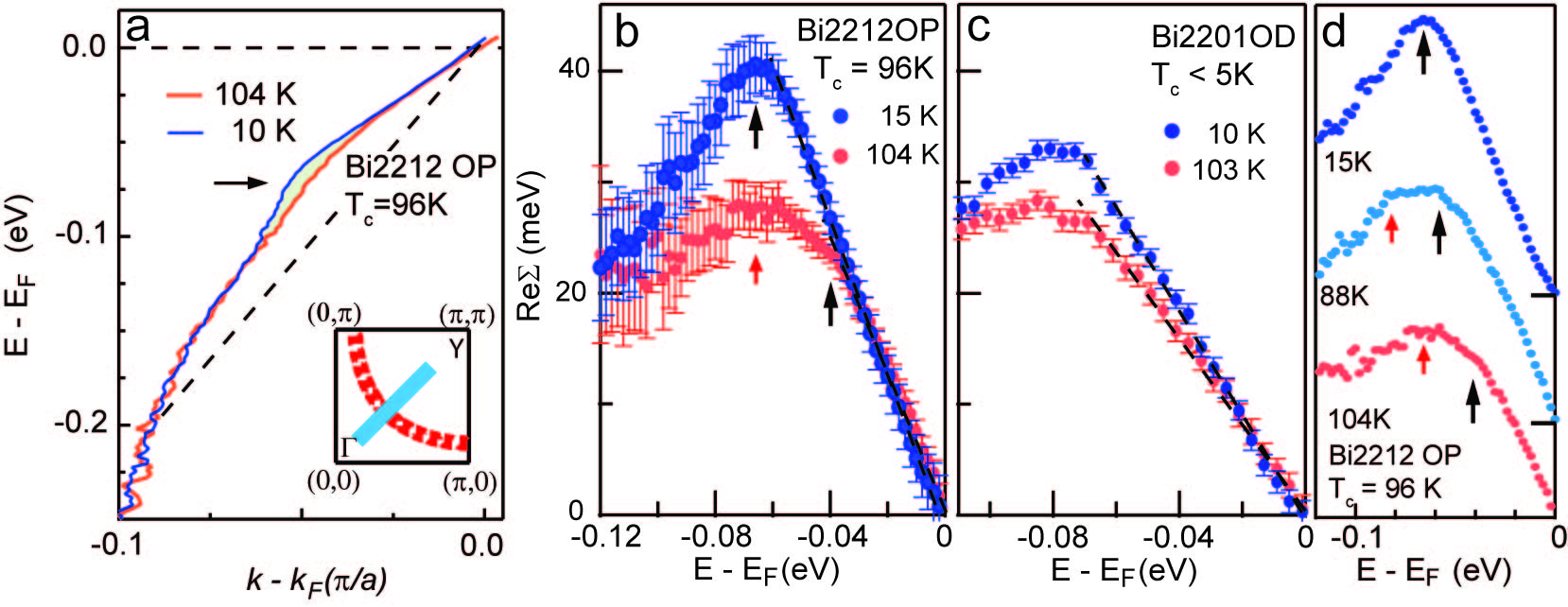}
\centering
\caption{\label{Fig 12: Temperature dependence of nodal dispersion} (a) Optimally-doped Bi-2212: nodal cut at 10K (SC) and 104K (non-SC).  Dashed line denotes assumed linear bare band.  The dominant kink is near 70meV (arrow) for at both temperatures.  (b) Re$\Sigma$ at 15K and 104K, approximated by subtracting linear bare band in (a) from dispersion.  At 15K, there is one dominant feature near 70 meV, and at 104K there is an additional shoulder or subkink near 40 meV. (c) Re$\Sigma$ extracted in the same way for OD Bi-2201 (T$_c$$<$5K), which is non-SC at both 10K and 103K.  Position of features is the same at both temperatures, suggesting that the temperature dependence in (b) is connected to the SC transition. (d) OP Bi-2212: Re$\Sigma$ at three temperatures, demonstrating the shifting of the subkink across the SC transition.}
\end{figure}

Above T$_c$, the gap is closed on part of the Fermi surface, so a downshifting of the kink energy is expected.  Fig. \ref{Fig 12: Temperature dependence of nodal dispersion}(a) shows the nodal band dispersion above and well below T$_c$ in optimally doped Bi-2212, which is found by standard momentum distribution curve (MDC) analysis,\cite{SelfEnergyEffectsPhotoemission:Johnson} and Re$\Sigma$ is approximated by subtracting an assumed linear bare band from the measured band dispersion. Re$\Sigma$ at both 15K and 104K, shown in Fig. \ref{Fig 12: Temperature dependence of nodal dispersion}(b), has a pronounced peak near 70 meV, consistent with previous findings that the apparent kink energy does not change across T$_c$.\cite{ElectronPhononCoupling:Lanzara, Gromko:Eboson_couple}  This lack of gap-shifting across T$_c$ complicated the explanation of this kink, but we have shown that this issue can be clarified via a temperature-dependence study where additional features become apparent at elevated temperatures.\cite{MultipleKinks:Lee}  A closer look at Re$\Sigma$ at 104K reveals a shoulder--a subkink-- near 40 meV, suggesting that an additional energy scale comes into play at elevated temperature.  This peculiar temperature dependence cannot be attributed to a thermal effect, as overdoped Bi-2201 with T$_c$ $<$ 5K shows no such change between similar temperatures. Thus the appearance of the subkink is more readily attributed to the loss of superconductivity between 15K and 104K. In the absence of superconductivity, the energies of the modes can be identified.  The mode at 70 meV agrees with the energy of the oxygen half-breathing mode observed by neutron scattering,\cite{McQueeney:NeutronHalfBreathing} and the mode at 40meV agrees in energy with the B$_{1g}$ buckling phonon.\cite{Cuk:B1gAN}   Studying an intermediate temperature can verify this assignment, as the near-nodal gap in optimally doped Bi-2212 has a pronounced temperature dependence\cite{Lee:twoGapARPES_TDep}.  As shown in Fig. \ref{Fig 12: Temperature dependence of nodal dispersion} (d), Re$\Sigma$ at 88K has two pronounced features at approximately 60 meV and 90 meV, consistent with partial gap-shifting of the phonons associated with the normal-state kink.  The temperature dependence of the relative strength of these modes can be captured by a Holstein model, as discussed in Ref. \cite{MultipleKinks:Lee}.

\begin{figure} [t]
\includegraphics [type=jpg,ext=.jpg,read=.jpg,clip, width=4.7 in]{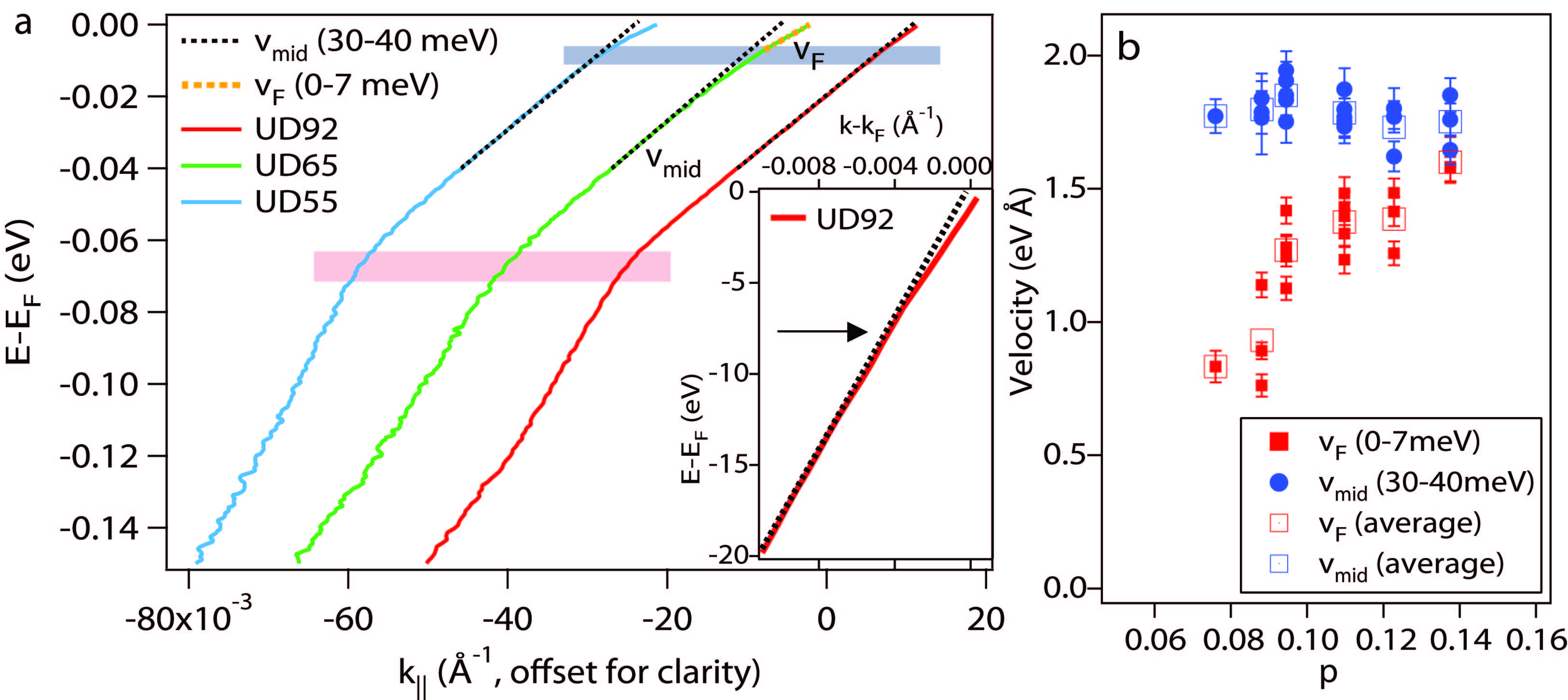}
\centering
\caption{\label{Fig 13: Low-Energy Kink} (a) MDC-derived band dispersions along the nodal direction for three dopings of Bi-2212 measured by laser-ARPES: UD55, UD63, and UD92, offset horizontally for clarity.  Black dotted line denotes velocity fit between 30-40 meV (v$_{mid}$).  The velocity at E$_F$ (v$_F$, 0-7 meV, orange dashed line accompanying UD63 dispersion) is smaller than v$_{mid}$, indicating a low-energy renormalization. Grey and pink bars denote approximate positions of low-energy kink and 70 meV kink respectively.  Inset: detail of UD92 dispersion below 20meV. (b) v$_F$ and v$_{mid}$ as a function of doping, demonstrating doping-dependent v$_F$ and an increasing renormalization with underdoping.  Open symbols indicate the average velocities for each doping.}
\end{figure}

In addition to the multiple-mode coupling near 70 meV, renormalization effects at much lower energy scales have also been observed following the advent of high-resolution laser-based ARPES.\cite{Liu:HighResLaserARPES}  A low-energy kink ($<$10 meV) has been reported in the nodal dispersion of optimally-doped Bi-2212.\cite{LaserARPES:LEKinkPlumb,Rameau:LEKink,Vishik:LEKink}  Other papers have reported a decrease of the MDC linewidth at similar energy.\cite{Bi2212NewCouplingLaser:Zhang}  Notably, this energy-scale is smaller than the superconducting gap, suggesting somewhat different physics than the 70 meV kink.  Recently, we have studied the doping-dependent systematics of the low-energy kink,\cite{Vishik:LEKink} and the nodal dispersions for three dopings (UD55, UD63, and UD92) are shown in Fig. \ref{Fig 13: Low-Energy Kink} (a).  The velocity at the Fermi level (v$_F$, linear fit 0-7 meV) is smaller than that between 30-40 meV (v$_{mid}$), indicating the presence of a kink at some intervening energy.  Notably, instrumental energy resolution and thermal broadening would push the near-E$_F$ velocity to higher values,\cite{LaserARPES:LEKinkPlumb} so this apparent slowing of charge carriers close to E$_F$ is not captured by trivial effects.  Moreover, a low-energy feature is observed both in the MDC dispersion and width--which are related to Re$\Sigma$ and Im$\Sigma$ respectively--and the presence of the kink at at least three different dopings (Fig. \ref{Fig 13: Low-Energy Kink} (a)), strongly suggests that it is a ubiquitous aspect of the nodal physics in Bi-2212.  It is then important to modify an earlier result--the universal nodal Fermi velocity--which was reported before ARPES resolution was sufficient to resolve the low-energy kink.\cite{UniversalNodalvF}  In Fig. \ref{Fig 13: Low-Energy Kink}(b) we plot v$_F$ and v$_{mid}$ as a function of doping.  The latter is largely independent of doping, but the former decreases substantially with underdoping, tracking the superconducting dome and showing an increasing localization of nodal quasiparticles as hole concentration is decreased.  Taken together with measurements of the near-nodal gap,\cite{Tanaka:twoGapARPES_dopingDep} this phenomenology has immediate implications for the interpretation of thermodynamic data, but the origin of this low-energy kink--and whether its doping-dependence arises from the same physics which leads to the diminishing of superconductivity with underdoping--is yet to be explained.

Along the node, low-energy excitations aside from superconductivity can be studied, and renormalization effects at multiple energies are revealed.  Although a single kink is seen near 70 meV, a closer analysis reveals temperature-dependent subkinks.  In addition, zooming in close to E$_F$ with high-resolution laser ARPES reveals a new low-energy kink, whose doping-dependence leads to a doping dependent v$_F$.  Understanding the origin of the multiple kinks, their doping and temperature dependencies, and their relation to superconductivity remain as challenges to the field.

\section{Conclusions}
We have presented a review on low-energy excitations in the cuprates, which gives evidence for the distinct nature of the pseudogap and superconductivity, and emphasized the importance of multiple bosonic modes in the nodal spectrum.  Evidence for a 'two-gap' scenario includes an increasing deviation of the superconducting gap function from a simple \textit{d}-wave form in the underdoped regime and different temperature dependencies of the gap in different regions of momentum space.  We showed Bi-2201 data which presented a striking case where the phenomenology of the pseudogap state is consistent with a fluctuating density-wave order, though it remains to be seen whether this density-wave picture is robust for other cuprates.  Though there is evidence that the 'normal' state of the cuprates is distinct from superconductivity, we are still left with the task of uncovering the identity of that state and understanding if and how it promotes high temperature superconductivity.  Likewise, we have a great deal of experimental phenomenology about the 70meV kink and the low-energy kink--pointing to a multiple-phonon explanation for the former and deriving a doping-dependent v$_F$ from the latter--but the link between these kinks and high-temperature superconductivity, if any, remains an unsolved puzzle.

\begin{acknowledgements}
SSRL is operated by the DOE Office of Basic Energy Science. This work is supported by DOE Office of Basic Energy Science,
Division of Materials Science, with contracts DE-FG03-01ER45929-A001 and DE-AC02-76SF00515.
\end{acknowledgements}

\bibliography{CuprateReview}
\end{document}